\documentstyle[12pt]{article}
\title{Boltzmann and Statistical Mechanics $^\dagger$}
\author{E. G. D. Cohen\\
The Rockefeller University\\
New York, NY 10021, USA}
\date{  }
\begin{document}
\maketitle

\vspace{.50in}

``O! immodest mortal!  Your destiny is the joy of watching the 
evershifting battle!''
\begin{flushright}
Ludwig Boltzmann
\end{flushright}

\vfill

$^\dagger$ Lecture given at the International Meeting ``Boltzmann's Legacy -
150 Years after his Birth'', organized by the Accademia Nazionale
dei Lincei, 25 - 28 May 1994, in Rome, to be published in: ``Atti dell"Accademia
Nazionale dei Lincei'', 1997.

\newpage
\hsize=6in
\hoffset=-.5in
\baselineskip=1.5\baselineskip
\section{Introduction}

I received two invitations to this meeting: in the first unofficial one I was
asked to speak "on the transport properties of dense gases".  This involves
the generalization of the Boltzmann equation to higher densities, a topic
on which I have worked for more than 35 years.  Later, I also received an 
official invitation, in which I was asked to give a lecture of ``a generalized
character''.  Although the first topic would be a natural and relatively
easy one, since I have spoken on it often and thought about it a lot, the
second one seemed much more difficult but irresistibly challenging, in
allowing me to view Boltzmann's work in the last century from the
perspective of the end of this century.  This seems at first sight to be a 
precarious undertaking for a research
scientist, but, as I hope to make clear to you, there may be advantages to
this.  While the historian of science is able to place the work of 
a scientist of the past in the context of that of his contemporaries the 
research scientist can place the work of that scientist in the
context of present day research and, up to a point, identify with his
difficulties and achievements in the past on the basis of his own experience
in the present day.  I embark then on my
perilous self-imposed task in the hope of providing some new
perspectives on Boltzmann and his work, which are, I hope,  historically not 
too inaccurate as far as the past is concerned, and stimulating, if not
provocative, as far as the future is concerned.

\section{Boltzmann, mechanics and statistics.}

At the time Boltzmann, born in 1844, began his career, Mechanics was the 
queen of
theoretical physics, by far the most completely developed part of
theoretical physics, the example as well as the ultimate goal for all other
branches of theoretical physics.  In the second half of the 19th century
two major obstacles were to present themselves to this: the Second Law
of Thermodynamics and Electromagnetism.   Boltzmann's first attempt to
``mechanize'' the Second Law can be found already in his second paper, 
published at age 22 in 1866, entitled ``On the Mechanical Meaning of the 
Second Law 
of the Theory of Heat''$^{[1]}$.  It is good to keep in mind that the Second 
Law consists of two parts; 1. a reversible and 2. an irreversible one.  1.
introduces the existence of an integrating factor, the inverse absolute
temperature 1/T, for the heat dQ reversibly supplied to or removed from a
system, such that dQ/T = dS, the (total) differential of the entropy S; 2.
states that the entropy of an isolated (adiabatic) system can never
decrease. In section IV of Boltzmann's 1866 paper:  ``Proof of the Second
Law of the Mechanical Theory of Heat'' he mainly addresses the first
aspect and is very cavalier about the second, the more difficult or,
perhaps better, intractable one.  He first deals with the case that heat is
supplied to a system under the condition of equality of the inner and outer
pressure of the system, and he shows that in that case $\oint$ dQ/T = $\oint$
dS = 0. He then argues that if this equality of pressures does not obtain, dQ 
must be smaller, so that in that case $\oint$ dQ/T $<$ 0.  This is clearly at 
best a physical argument not a mechanical proof!
	
Two years later in a paper called: ``Studies on the Equilibrium of the
Kinetic Energy Between Moving Material Points''$^{[2]}$, he follows Maxwell in
introducing probability concepts into his mechanical considerations and
discusses a generalization of Maxwell's distribution function for point 
particles in free space to the very
general case that ``a number of material points move under the influence
of forces for which a potential function exists.  One has to find the
probability that each one of them moves through a given volume with a
given velocity and velocity direction''.  This is the first of many papers in
which Boltzmann discusses and generalizes Maxwell's velocity distribution for 
point particles in free space to the case that external forces are 
present$^{[2,3]}$ and to (polyatomic) molecules$^{[4]}$, leading to the
Boltzmann factor and the Maxwell-Boltzmann distribution function$^{[5]}$.

Yet in 1872, when Boltzmann derived in his paper: ``Further Studies on
Thermal Equilibrium Between Gas Molecules'' what we now call the
Boltzmann equation for the single particle position and velocity
distribution function in a dilute gas$^{[6]}$ and used, following Clausius and
Maxwell, what the Ehrenfests called the Stoszzahl Ansatz, he does not seem
to have fully realized the statistical nature of this assumption and
therefore also of the ensuing H-theorem, for the approach to equilibrium. 
He says$^{[7]}$: ``One has therefore rigorously proved that, whatever the
distribution of the kinetic energy at the initial time might have been, it
will, after a very long time, always necessarily approach that found by
Maxwell''.

The beauty of the H-theorem  was that it derived in one swoop both
aspects of the Second Law:  first the (irreversible) approach to thermal
equilibrium and then, from the value of the H-function {\em{in}} equilibrium, 
the connection between the H-function and Clausius' entropy S: H = --const. 
S + const.  Only later forced by Loschmidt's Reversibility Paradox$^{[8]}$ and
Zermelo's Recurrence Paradox$^{[9]}$, as the Ehrenfests were to call 
them$^{[10]}$, did Boltzmann clearly state the probabilistic nature of the 
Stoszzahl Ansatz, viz. that the Stoszzahl Ansatz and the H-theorem only held for
disordered states of the gas and that these states were much more
probable than the ordered ones, since the number of the first far exceeded
that of the second.

In the paper itself, however, this is never mentioned; it is as if the
Ansatz was self evident.  Therefore, Boltzmann did {\em{not}} derive here the
Second Law purely from mechanics alone either and till the present day,
no mechanical derivation of the Boltzmann equation exists, although the
Stoszzahl Ansatz must ultimately be derivable from the mechanics of a
very large number N of particles, i.e., from ``large N-dynamics''.

I must admit that I find it difficult to assess Boltzmann's precise
attitude towards the mixture of mechanics and statistics that a
description of the behavior of macroscopic systems -- gases mainly for
him -- necessitates.  Uhlenbeck, a student of Ehrenfest's, who was himself
a student of Boltzmann's, told me several times:  ``Boltzmann was
sometimes confusing in his writings on the statistical aspects of his
work and this, in part, prompted the Ehrenfests to write their
clarifying and in a way definitive article to answer his opponents$^{[10]}$''.

The depth of ill-feelings generated by Boltzmann's exhausting
discussions with his German colleagues, especially the Energeticists$^{[11]}$ 
and the resistance to his ideas, in particular with regards to the H-theorem
that surrounded him, still resonated for me when Uhlenbeck said to me one
day in some mixture of anger and indignation:  ``that damned Zermelo, a 
student of
Planck's, nota bene''$^{[12]}$, an echo after two generations of past injustice
and pain inflicted on Boltzmann by his hostile environment.   Let me quote
Boltzmann himself in his introduction to his response to Zermelo in 1896,
for another aspect of his isolation and near desperation.  After having
explained that he has repeatedly and as clearly as possible emphasized in
his publications that the Maxwell distribution function as well as the
Second Law are of a statistical nature, he says$^{[9a]}$:  ``Although the
treatise by Mr. Zermelo ``On a Dynamical Theorem and the Mechanical
Theory of Heat'' admittedly shows that my above mentioned papers have
still not been understood, nevertheless I have to be pleased with his
article as being the first proof that these papers have been noticed in
Germany at all''.

I want to cite a second indication of this solitude. Boltzmann wrote 
a letter to H. A. Lorentz in 1891, in response to a letter Lorentz
sent to him, which pointed out for the second time an error in one of his
papers$^{[13]}$: ``Already from the postmark and the handwriting I knew that
the letter came from you and it pleased me.  Of course, each letter from
you implies that I have made an error; but I learn then always so much,
that I would almost wish to make still more errors to receive even more
letters from you''.  This quotation must be seen in the above mentioned
context of Boltzmann's isolation in the German speaking countries, since in
an earlier letter to Lorentz in 1886, in response to the above mentioned
earlier error he made, he says$^{[14]}$:  ``I am very pleased that I have 
found in you someone who works on the extension of my ideas about the theory 
of gases.  In Germany, there is almost no one who understands this properly''.

Probably motivated by the opposition of his contemporaries to his
mechanical or kinetic method to prove the Second Law on the basis of the
Boltzmann equation via the H-theorem, Boltzmann switched completely in
1877, when he introduced his statistical method, as the Ehrenfests called
it, with no mechanical component in it at all, leading to the famous
relation between entropy and probability.  In this 1877 paper:  ``On the
Relation Between the Second Law of Thermodynamics and Probability
Theory with Respect to the Laws of Thermal Equilibrium$^{[15]}$'' he begins by
saying$^{[16]}$: ``A relation between the Second Law and probability theory
showed, as I proved$^{[4b]}$, that an analytical proof of it is possible on no
other basis than one taken from probability theory''.

Boltzmann emphasizes this necessity of probability concepts to understand the 
Second Law throughout the 1890's, when he mainly argued
with his opponents over the interpretation of the H-theorem:  his
creative period had lasted about twenty years and one could ask to what
extent this had been influenced by his difficulties with his contemporaries. 
To be sure, Boltzmann's explanations of the crucial points concerning the 
interplay between mechanics and probability which were at the heart of 
Loschmidt's and Zermelo's objections to the H-theorem, although basically 
correct, did not capture all the subtleties of the necessary arguments and 
together with the hostile Zeitgeist made his efforts largely unsuccessful, 
certainly for himself.  One must admit, though, that even today it is not 
easy to explain the paradoxes clearly, even to a sympathetic audience!

In particular, in his rebuttal of Zermelo's Recurrence Paradox$^{[9]}$ in
1896 and 1897 Boltzmann argued as clearly as he could, from a great
variety of points of view, just as he had done twenty years earlier against
Loschmidt's Reversibility Paradox$^{[8]}$.  It seems like a last vigorous
attempt to show once and for all that there was really no conflict
between mechanics and his kinetic theory.  It was to no avail and this
must have greatly depressed him.  When Einstein in 1905 proved the
existence of atoms by Brownian motion$^{[17]}$, it was far too late: Einstein 
was still unknown in 1905 and Boltzmann had probably given 
up long before then. I do not know whether Boltzmann ever read or heard of 
Einstein's paper, but if he did, although other causes undoubtedly played a 
role, it might  -- considering the state he must have been in -- have 
contributed to, rather than prevented, his suicide in 1906. A systematic, 
critical and very structured account of Boltzmann's arguments was finally 
presented in 1909 - 1911 by the Ehrenfests' ``Apologia'' in their above 
mentioned Encyclopedia article$^{[10]}$.

It is ironic perhaps to note that Boltzmann's second approach, the
statistical method, introducing what we now call Boltzmann statistics,
has been in retrospect much more influential than the first, the kinetic
method.  This is in part because it has turned out that a meaningful 
generalization of Boltzmann's equation to higher densities, as well as 
obtaining concrete results from such an equation, have proved very difficult.
Nevertheless, many new deeper insights into the behavior of dense 
nonequilibrium fluids have resulted from this work$^{[18]}$.  Boltzmann 
himself clearly preferred the kinetic method over the statistical method, 
because it was based on the dynamics, i.e., the collisions between the 
molecules out of which the gas consists and therefore allowed a direct 
connection with the motion of the particles.  That statistics also came in 
was finally due to the presence of very many particles, but no substitute 
for the basic mechanical nature of the behavior of gases.  As if he had a 
premonition of this future development, Boltzmann's summarizing ``Lectures 
on Gas Theory''$^{[19]}$ are almost exclusively devoted to the kinetic method 
and hardly mention the statistical method at all.

Boltzmann never lost his predilection for mechanics.  In his 1891-1893 lectures
on Maxwell's theory of electromagnetism$^{[20a]}$, he used, wherever he could,
elaborate mechanical analogies, by endowing the ether with all kinds of 
intricate mechanical properties.  This had been started by Maxwell himself in 
1856$^{[21]}$ but later Maxwell abandoned this approach in his presentations. 
Boltzmann was very well aware of his ``old-fashioned'' mechanical predilection. 
I quote what he said, in a lecture ``On Recent Developments of the Methods of 
Theoretical Physics'', at a Naturforschung meeting in Munich in 1899$^{[22]}$. 
After having described the situation in theoretical physics as it existed at 
the beginning of his studies when$^{[23]}$ ``the task of physics seemed to 
reduce itself, for the entire future, to determining the force between any 
two atoms and then integrating the equations that follow from all these 
interactions for the relevant initial conditions,'' he continues: ``How 
everything has changed since then!  Indeed, when I look back at all these 
developments and revolutions, I see myself as an old man in scientific 
experience.  Yes, I could say, I am the only one left of those who still 
embrace the old wholeheartedly, at least I am the only one who still fights 
for it as much as he can ... I present myself therefore to you as a reactionary,
a straggler, who adores the old, classical rather than the newer things...''.

Although it was Boltzmann himself who introduced the idea of an ensemble of 
systems (he called it ``Inbegriff'' or ``collection''), this (probabilistic) 
``trick'', as he later called it, did not deter him from a deep mechanical 
point of view as well.  The ensemble is first mentioned in the beginning of a 
paper in 1871 entitled$^{[4b]}$ ``Some General Theorems on Thermal 
Equilibrium'': ``One has a very large number of systems of material points 
(similar to a gas which consists of very many molecules, each of which itself 
is again a system of material points).  Let the state of any one of these 
point systems at any time $t$ be determined by $n$ variables $s_1...s_n$; ........
we have only to assume that
between the material points of the various point systems no interaction
ever occurs.  What one calls in the theory of gases the collisions of the
molecules, will be excluded in the present investigation......The number of
variables $s$ that determines the state as well as the differential equations
[of motion] should be the same for all systems.  The initial values of the
variables $s$ and consequently the states at an arbitrary time $t$ on the other
hand should be different for the various point systems''$^{[24]}$.  It was 
a neat trick to compute the macroscopic observables of a gas as an average over
many samples of the gas, each with different initial coordinates and
momenta of the gas particles, assuming tacitly equal a priori probability
of all the possible microstates of the gas with the same total energy.  It
was much easier than following the motion of all the particles in a given
system in time and then taking a time average.  Boltzmann used it to
determine the equilibrium distribution function for a gas in thermal
equilibrium.  Curiously enough, Maxwell used the same idea independently
in an 1879 paper entitled$^{[25]}$:  ``On Boltzmann's Theorem on the Average
Distribution of Energy in a System of Material Points''.  He says$^{[26]}$: 
``I have found it convenient, instead of considering one system of material
particles, to consider a large number of systems similar to each other in
all respects, except in the initial circumstances of the motion, which are
supposed to vary from system to system, the total energy being the same
in all.  In the statistical investigation of the motion, we confine our
attention to the {\em{number}} of those systems which at a given time are in a
phase such that the variables which define it lie within given limits". 
Maxwell does not mention Boltzmann here because he probably stopped 
reading Boltzmann after 1868, due to the - for him -
excruciating amount of detail in the latter's papers$^{[27]}$.

That Boltzmann had indeed not abandoned his hopes of giving a mechanical
interpretation of the Second Law is borne out by his work on
Helmholtz's monocycles, first published in 1884, in his paper$^{[28]}$:  
``On the Properties of Monocyclic and Other Related Systems'', which allowed a
formal analogy between appropriate changes of these simple mechanical
systems characterized by a single frequency$^{[29]}$ and those appearing in
the First and the Second Law for reversible thermodynamic changes.  He
begins this paper as follows$^{[30]}$:  ``The most complete mechanical proof of
the Second Law would clearly consist in showing that for each arbitrary
mechanical process equations exist that are analogous to those of the
Theory of Heat.  Since, however, on the one hand the Law does not {\em{seem}}
[my italics] to be correct in this generality and on the other hand, because
of our ignorance of the nature of the {\em{so-called}} [my italics] atoms, the
mechanical conditions under which the heat motion proceeds cannot be
precisely specified, the problem arises to investigate in which cases and
to what extent the equations of mechanics are analogous to those of the
Theory of Heat''.  He continues:  ``One will not be concerned here with the
construction of mechanical systems, which are completely identical with
warm bodies, but rather with identifying all systems, that exhibit behavior
more or less analogous to that of warm bodies''.  He further elaborates on
this analogy in two more papers in the following years$^{[31]}$.  I quote the
beginning of the third paper$^{[31b]}$ ``New Proof of a Theorem Formulated by
Helmholtz Concerning the Properties of Monocyclic Systems'', to illustrate
the importance Boltzmann attached to this work:  ``The great importance
which the introduction of the notion of monocyclic systems and the
development of their most important properties has for all investigations
concerning the Second Law should make the following considerations
appear as not completely superfluous...''.

As an aside, I remark that Boltzmann's writings on the existence of
atoms seem to be ambivalent and a mixture of, on the one hand, actually
using atomism all through his works and elaborately discussing the many
arguments in favor of it$^{[32a]}$, while, on the other hand, stating in writing
the possibility of other equally valid descriptions of nature, as provided
by the Energeticists or phenomenologists.  Thus in his 1899 lecture 
mentioned above, he says$^{[32b]}$: ``From this follows
that it cannot be our task to find an absolutely correct theory, but rather the
simplest possible picture which represents experiment as best as possible. 
One could even think of the possibility of two entirely different theories,
which are both equally simple and agree with the phenomena equally well,
which therefore, although completely different, are both equally correct''. 
Although true, I find it hard to escape the impression that this statement
was meant more as an attempt to assuage his Energeticists opponents than as an 
account of
his actual position.  The difficulty, of course, was that at the time 
evidence for the existence of atoms was ultimately only circumstantial and 
that no direct experimental demonstration in any fashion had yet been given.

In the same 1884 paper mentioned above, Boltzmann introduces the
notion of Ergoden$^{[33]}$.  It was used by Boltzmann as an equilibrium
ensemble of systems (Gibbs' micro-canonical ensemble).  According to the
Ehrenfests$^{[10]}$, Boltzmann defines an ergodic system as one whose
unperturbed motion goes, when indefinitely continued, finally {\em{``through
each phase point''}} that is consistent with its given total energy.  In this
way Boltzmann suggested how to understand on the basis of the dynamics
of the gas molecules, i.e., from mechanics, that ``ergodic (ensemble)
averages''$^{[34]}$ in phase space could replace the time averages through
which the macroscopic properties of the gas were defined.  In their
article, the Ehrenfests argued that Boltzmann's requirement for an
ergodic system was too strong and they replaced it by introducing a quasi-
ergodic system ``which approaches each point of the energy surface
arbitrarily close''.  The idea is that under such conditions the time and
ensemble averages would still be the same.  The classical ergodic theory
culminated in Birkhoff's ergodic theorems$^{[35]}$ proving the existence of the
time average and then for metrically transitive systems the equality of
time and phase space averages.  Later, it was the application of dynamical
systems theory and the introduction of the concept of measure which
allowed the more precise unification of mechanics and statistics that
Boltzmann had in mind. 	

As in electromagnetic theory, so in the theory of gases, the
mechanical aspects have been obliterated.  In the theory of gases this
occurred not only through Boltzmann's statistical method but mainly
through Gibbs' 1902 book ``Elementary Principles in Statistical
Mechanics''$^{[36]}$.  It was also there that the term ``Statistical Mechanics''
was first introduced.  Here, on purpose, all reference to the molecular
"constitution of matter" was as much as possible avoided to achieve a
generality similar to that of thermodynamics.  The book was mainly devoted to
a study of thermal equilibrium:  the canonical ensemble, already
introduced by Boltzmann$^{[29]}$ for a system in thermal equilibrium at a
given temperature, as well as the micro-canonical ensemble were given
their names and extensively studied.  Their connection with
thermodynamics was made via a thermodynamic analogy$^{[37]}$, as had been
done earlier by Boltzmann$^{[4c,15,28]}$.  While thermodynamics never changed
and was essentially unaffected by the advent of quantum mechanics -
except for the addition of the third law, which is not as absolute and
general as the first two - Gibbs' ensembles had only to be modified
slightly to accommodate quantum mechanics.

The Ehrenfests were rather critical of Gibbs' contributions to
statistical mechanics and they gave a very subjective presentation of his
accomplishments$^{[38]}$.  The main criticism was that Gibbs had only devoted
essentially one chapter - and a purely descriptive at that - to the problem
of the approach to equilibrium (ch.12), a problem that was central in the
considerations of Boltzmann and the Ehrenfests.  In this chapter Gibbs
describes a generalization of Boltzmann's H-theorem in $\mu$-space, the
phase space of one molecule, to $\Gamma$-space, the phase space of the entire 
gas. This consideration, as well as the rest of the book, was entirely based on
probability notions, and the whole basic molecular mechanism of
collisions - viz. that of binary collisions for the dilute gas which
Boltzmann had considered - was completely absent.  I think the
Ehrenfests' critical attitude in this has turned out to be unjustified since
Gibbs' statistical mechanics has been far more influential than they
surmised at the time.  In fact, it has dominated the entire twentieth
century and only now, with a renewed interest in nonequilibrium
phenomena, is a revival of Boltzmann's mechanistic approach reemerging. 
Here one should keep in mind two things.  First, even simple
nonequilibrium phenomena are often far more difficult to treat than many
rather complicated equilibrium phenomena.  Second, starting with L. S.
Ornstein's Ph.D. thesis ``Application of the Statistical Mechanics of Gibbs
to Molecular-Theoretical Questions'' written under Lorentz's direction 
in 1908$^{[39]}$, the
calculation of the thermodynamic properties of a gas in thermal
equilibrium via the canonical and related distribution functions turned out
to be far simpler than those based on the microcanonical ensemble.  Gibbs'
statistical mechanics has not only led to enormous advances in
equilibrium statistical mechanics but also, by virtue of, for instance, the
introduction of the Renormalization Group in the theory of critical
phenomena, to entirely new ways of thinking about what is relevant in
nonequilibrium statistical mechanics.  And yet ...\\
\section{Return of Mechanics - Boltzmann's Heritage}

The revival of the role of mechanics in statistical mechanics is due
to important new developments in mechanics itself or as it is now called
the theory of dynamical systems, especially by the Russian school of
mathematics, emanating from A. N. Kolmogorov and Ya. G. Sinai.  This has 
lead in recent
years to a beginning of a dynamical formulation of statistical mechanical
problems, especially for nonequilibrium systems, near or far from
equilibrium.  Far from equilibrium means here a system with large
gradients and therefore large deviations from Maxwell's (local)
equilibrium velocity distribution function, where hydrodynamics cannot be
applied, not a turbulent system, where a hydrodynamic description is still
(believed to be) applicable.  I will quote three examples.

1. Boltzmann's notion of ergodicity for Hamiltonian systems in
equilibrium on the energy surface has, in a way, been generalized by
Sinai, Ruelle and Bowen (SRB)$^{[40]}$ to dissipative systems in a
nonequilibrium stationary state.  In that case the attractor, corresponding
to the nonequilibrium stationary state, plays the role of the energy
surface in equilibrium and the SRB measure on the attractor in terms of
expanding, i.e., positive Lyapunov exponents (which determine the rate of
exponential separation of two initially very close trajectories) replaces
the Liouville measure on the energy surface for systems in equilibrium. 
This allows the assignment of purely dynamical weights to a macroscopic system
even far from equilibrium, since the molecules, whose dynamical
properties one uses, do not know how far the system is from equilibrium,
as equilibrium is a macroscopic non-molecular concept, or, to paraphrase
Maxwell$^{[41]}$: ``When one gets to the molecules the distinction between
heat and work disappears, because both are [ultimately molecular]
energy''.  This approach seems to differ in principle from the conventional
ones, based on extending Gibbs' equilibrium ensembles to nonequilibrium,
e.g. via a Chapman-Enskog-like solution of the Liouville equation for the
entire system (instead of for the Boltzmann equation, for which it was
originally designed).  To be sure, these Gibbsian nonequilibrium ensembles
also contain dynamics - no nonequilibrium description is possible without
it - but not as unadulterated as in the SRB-measure.  The farthest one has
gone with these nonequilibrium Gibbs ensembles is the Kawasaki
distribution function$^{[42]}$ and it is not excluded that there is an intimate
connection between this distribution function and the SRB measure.

Recently the SRB measure has been checked for the first time far
from equilibrium for a many (56) particle system by a computer
experiment$^{[43]}$.  Here one studies very large temporary fluctuations of a
shearing fluid in stationary states with very large shear rates.  The ratio
of the fluctuations of the stress tensor to have, during a finite time, a
given value parallel or opposite to the applied shear stress, i.e. consistent
with or ``in violation of''  the Second Law, respectively, was measured and
found to be given correctly on the basis of the SRB measure.  In fact, one
has recently been able to indicate how this, in a way typical
nonequilibrium statistical mechanical system, can perhaps be discussed
on the basis of the SRB measure:  at least a scenario has been formulated,
where one could hope, at least in principle, to derive the just mentioned
result rigorously from the SRB measure  with ``large N-dynamics''$^{[44]}$. 

2. Relations between the transport coefficients and the Lyapunov
exponents of a fluid in a nonequilibrium stationary state have been
uncovered.  While the former refer to hydrodynamic, i.e.,
nonequilibrium properties in ordinary three dimensional space, the latter
refer to the dynamical behavior on the attractor in the multidimensional
phase-space of the entire system.  For the above mentioned many particle
shearing fluid in a nonequilibrium stationary state, one has obtained an
explicit expression for the viscosity coefficient of this fluid in terms of
its two maximal - i.e., its largest and its smallest - Lyapunov
exponents$^{[45]}$.  This expression gives a value for the viscosity which
agrees numerically with that found directly from computer simulations
for a 108 and 864 particle shearing fluid, not only in the linear
(hydrodynamic) regime near equilibrium, where the viscosity is
independent of the imposed shear rate, but even in the non-linear
(rheological) regime, far from equilibrium, where the viscosity
coefficient itself depends on the shear rate.

3. A somewhat analogous expression has been derived for the linear
diffusion coefficient of a point particle in a regular triangular array of
hard disks$^{[46]}$.  In addition, the diffusion coefficient$^{[47,48a]}$ 
as well as the pressure$^{[48b]}$ for this system have been computed using a 
cycle-expansion, i.e., an expansion in terms of the periods and Lyapunov
exponents of the unstable periodic orbits of the particle in the hard disk
system.  Furthermore a number of rigorous results have been proved for
the linear transport behavior of this system based on the SRB measure$^{[49]}$.

In all these cases the re-emergence of dynamics appears in a global
Gibbsian-sense, in that it involves global Lyapunov exponents of the entire
system, not detailed collision dynamics between small groups of particles
as in kinetic theory.

It appears therefore that Boltzmann's attachment to mechanics, as
exemplified by his unceasing attempts at a mechanical interpretation of
the macroscopic behavior of many particle systems, in particular of the
Second Law of thermodynamics, has re-emerged after a hundred years,
thanks to new developments in mechanics.  I note that these new
developments, apart from the above mentioned statistical mechanical
problems, have so far been applied mainly to simple (one-dimensional)
maps and few particle (i.e., few degrees of freedom) systems.  For the
connection with statistical mechanics it seems important to introduce
methods into these dynamical considerations which use explicitly the very
large number of degrees of freedom typical for statistical mechanical,
i.e., for macroscopic systems.  Such ``large N-dynamics'' could lead to a
``statistical'' dynamics that might be crucial to bridge the gap between
the prevailing rigorous treatments of dynamical systems of a few degrees
of freedom on the one hand and statistical mechanics on the other hand. 
For example, important distinctions in ``small N-dynamics'' as to the exact
number of conservation laws (or of Axiom-A systems)$^{[50]}$ might be less
relevant in the ``large N-dynamics'' for macroscopic systems.

A case in point would be to establish a connection between
dynamical system theory and Boltzmann's kinetic theory of dilute gases. 
One could then ask:  what is the connection between the linear viscosity of a 
dilute gas as given by the Boltzmann equation in terms of binary collision
dynamics (e.g. as related to an eigenvalue of the linear Boltzmann
collision operator$^{[51]}$) and the above mentioned expression for the
viscosity in terms of its two maximal Lyapunov exponents?

Before I end this mostly scientific presentation of Boltzmann's work
in statistical mechanics and its heritage, I would like to remark that I
have not just illustrated Boltzmann's work from a scientific but also from
time to time from a human or psychological point of view.  I believe that the
latter aspects are too often missing in the discussion of scientists,
especially those of the recent past.  The ``psychological'' remarks are
usually confined to anecdotes and occasional non-scientific comments. 
This is in contrast to what happens in the arts, where musicians, painters
and especially writers are critically ``psychoanalyzed'' as to their
behavior, their motivations and the connection between their personality
and their work.  In fact, this does exist to some degree for some
scientists of the remote past, e.g. for Newton$^{[52]}$.  Of all nineteenth
century scientists Boltzmann seems to be one of the most openly human
and deeply tragic, i.e., an obvious candidate for such an endeavor.  I hope
that this often missing human dimension in the discussion of scientists
and their work will be developed, although it is admittedly difficult to
find writers who have both the scientific and the human perception and
depth to do this$^{[53]}$.

I would like to conclude with two quotations, one about the work and
one about the man.  The first is from the obituary lecture by Lorentz, given
one year after Boltzmann's death$^{[54]}$.  This quotation is perhaps even more
applicable now than it was then and expresses beautifully Boltzmann's
message for the future when Lorentz says:  ``The old of which Boltzmann
speaks [see page 9 above] has in our days, thanks especially also to his own
efforts, flowered to new, strong life, and even though its appearance has
changed and will certainly often change in the course of time, we may yet
hope that it will never get lost for science''.

The other is from Boltzmann's 1899 lecture ``On Recent Developments of the
Methods of Theoretical Physics'', mentioned before.  It demonstrates that
Boltzmann's deep love for science transcended all his suffering in
practicing it.  Here, after discussing the many achievements of atomism
and the molecular theory of matter, which cannot at all be obtained by
just using macroscopic equations alone without any further microscopic 
foundations - as is done in phenomenology or energetics - he asks$^{[55]}$:
``Will the old mechanics with the old forces ... in its essence remain, or 
live on one day only in history ... superseded by entirely different notions? 
Will the essence of the present molecular theory, in spite of all
amplifications and modifications,
yet remain, or will one day an atomism totally different from
the present prevail, or will even, in spite of my proof$^{[56]}$, the notion 
of an absolute continuum prove to be the best picture?''  He concludes: ``Indeed
interesting questions!   One almost regrets to have to die long before they
are settled.  O! immodest mortal!  Your destiny is the joy of watching the
ever-shifting battle!''$^{[57]}$.

\noindent {\bf{Acknowledgement}} I have very much profited from various papers
of M. J. Klein and from the books ``The Kind of Motion We Call Heat'' by S. G.
Brush.  I am also indebted to my colleagues G. Gallavotti, J. R. Dorfman, 
A. J. Kox, L. Spruch, A. Pais and especially M. J. Klein for helpful remarks.  
I thank my secretary, S. Rhyne for preparing the manuscript and her as well 
as my daughter, A. M.
Cohen, for their help in translating Boltzmann's German and to the latter
also for her help with the text.

\newpage 
\noindent {\bf{References}}\\
1. L. Boltzmann:  ``\"{U}ber die mechanische Bedentung des Zweiten
Hauptsatzes der W\"{a}rmetheorie'', Wien. Ber.{\underline{53}}, 195-220 (1866);
Wissenschaftliche Abhandlungen (W.A.), F. Hasen\"{o}hrl, ed., Chelsea Publ. Co.,
NY (1968) Band I, p.9-33.\\
2. L. Boltzmann:  "Studien \"{u}ber das Gleichgewicht der lebendigen Kraft
zwischen bewegten Materiellen Punkten", Wien. Ber.{\underline{58}}, 517-560 
(1868); W.A. Band I, p.49-96; id., ``L\"{o}sung eines mechanisches Problems'',
Wien. Ber.{\underline{58}}, 1035-1044 (1868); W. A. Band I, p.97.\\
3. See also, L. Boltzmann: ``\"{U}ber das W\"{a}rmegleichgewicht van Gasen 
auf welche \"{a}uszere Kr\"{a}fte wirken'', Wien. Ber.{\underline{72}}, 
427-457 (1875); W.A. Band II, p.1-30.\\
4. L. Boltzmann:  (a) ``\"{U}ber das W\"{a}rmegleichgewicht zwischen
mehratomigen Gasmolekulen'', Wien. Ber.{\underline{63}}, 397-418 (1871); 
W.A. Band I, p.237-258; (b) id., ``Einige allgemeine S\"{a}tze \"{u}ber 
W\"{a}rmegleichgewicht'', Wien. Ber.{\underline{63}}, 679-711 (1871); W.A. 
Band I, p.259-287; (c) id., ``Analytischer Beweis des zweiten Hauptsatzes der 
mechanischen W\"{a}rmetheorie aus den S\"{a}tzen \"{u}ber das gleichgewicht 
der lebendigen Kraft'', Wien. Ber.{\underline{63}}, 712-732 (1871); W.A. Band 
I, p.288-308; (d) id., ``Neuer Beweis zweier S\"{a}tze \"{u}ber das 
W\"{a}rmegleichgewicht unter mehratomigen Gasmolek\"{u}len'', Wien. Ber.
{\underline{95}}, 153-164 (1887); W.A. Band III, p.272-282.\\
5. Of Boltsmann's about 140 scientific papers around 18 deal with the
Second Law and 16 deal with Maxwell's equilibrium distribution function
or both.\\
6. L. Boltzmann:  ``Weitere Studien \"{u}ber das W\"{a}rmegleichgewicht unter
Gasmolek\"{u}len", Wien. Ber.{\underline{66}}, 275-370 (1872); W.A. Band I, 
p.316-402.\\
7. See ref.6, W.A., Band I. p.345.\\
8. L. Boltzmann, ``Bemerkungen \"{u}ber einige Probleme der mechanischen
W\"{a}rmetheorie'', Wien. Ber.{\underline{74}}, 62-100 (1877), section II.\\
9. (a) L. Boltzmann: ``Entgegnung auf die W\"{a}rmetheoretischen
Betrachtungen des Hrn. E. Zermelo'', Wied. Ann.{\underline{57}}, 778-784 
(1896); W.A. Band III, p.567-578; (b) id. ``Zu Hrn. Zermelos Abhandling 
``\"{U}ber die mechanische Erkl\"{a}rung irreversibler Vorg\"{a}nge'''', 
Wied. Ann.{\underline{60}}, 392-398 (1897), W. A. Band III, p.579-586; (c) id.
``\"{U}ber einen mechanischen Satz Poincar\'{e}'s'', Wien. Ber.
{\underline{106}}, 12-20 (1897); W. A. BAnd III, p.587-595.\\
10. (a) P. and T. Ehrenfest, ``Begriffliche Grundlagen der statistischen
Auffassung in der Mechanik'',Enzycl. d. Mathem. Wiss. IV,2,II,Heft 6, p.3-90
(1912); p.30-32 or in P. Ehrenfest, {\em{Collected Papers}} (North-Holland,
Amsterdam) p.213-309 (1959); p.240-242; (b) English translation by M. J.
Moravcsik: P. and T. Ehrenfest, ``The Conceptual Foundations of the
Statistical Approach in Mechanics'', Cornell University Press, Ithaca, NY
(1959), p.20-22.\\
11. The Energeticists wanted to explain all natural phenomena on the basis
of energy alone, the most general ``substance'' present in the world.\\
12. ``note well''.\\
13. A. J. Kox: ``H. A. Lorentz's Contributions to Kinetic Gas Theory'',
Ann. Sci. {\underline{47}}, 591-606 (1990);p.602.\\
14. A. J. Kox, l.c. p.598.\\
15. L. Boltzmann:  ``\"{U}ber die Beziehung swischen dem zweiten Hauptsatz
der mechanischen W\"{a}rmetheorie und der Wahrscheinlichkeitsrechnung
respektive den S\"{a}tzen \"{u}ber des W\"{a}rmegleichgewicht'', Wien.
Ber.{\underline{76}, 373-435(1877); W. A. Band II, p.164-223.\\
16. Ref. 15,p.164.\\
17. Cf. A. Einstein, ``Autobiographical Notes'' in:  {\em{``Albert Einstein:
Philosopher - Scientist''}}, P. A. Schilp, ed., The Library of Living
Philosophers, Vol. VII, Evanston, IL, (1949) p.47-49.  See also:  B. Hoffman,
{\em{``Albert Einstein, Creator and Rebel''}}, New  Amer. Libr., NY (1972) p.58,59.\\
18. See, e.g., E. G. D. Cohen:  ``Fifty Years of Kinetic Theory'', Physica A 
{\underline{194}}, 229-257 (1993).\\
19. L. Boltzmann, ``Vorlesungen \"{u}ber Gastheorie'', 2 vols (Barth, Leizig,
1896, 1898); engl. transl. by S. G. Brush as ``Lectures on Gas Theory'', 
Univ. of Calif. Press, Berkeley (1964).\\
20. L. Boltzmann, ``Vorlesunger \"{u}ber Maxwell's Elektrizit\"{a}tstheorie''
(Aus den Mitteilungen des naturwissenschaftlichen Vereins in Graz. August
1873.) in: ``Popul\"{a}re Schriften'',second ed., Barth, Leizig (1919) 
p.11-24.\\
21. J. C. Maxwell, ``On Faraday's Lines of Force'', Trans. Cambr. Phil. Soc.
{\underline{10}}, 27-83 (1856); Scientific Papers 1, Cambridge, U. K., p.155.\\
22. L. Boltzmann: ``\"{U}ber die Entwicklung der Methoden der theoretischen
Physik in neuerer Zeit'', in: {\em{``Popul\"{a}re Schriften''}}, second ed., 
Barth, Leipzig (1919) p.198-227.\\
23. Ref. 22,p.204,205.\\
24. Ref. 4b, p. 259-260.\\
25. J. C. Maxwell, ``On Boltzmann's theorem on the average distribution of
energy in a system of material points'', Cambr. Phil. Soc. Trans.
{\underline{12}}, 547-575 (1879); Scientific Papers 2, Cambridge, U. K.,
p.713-741 (1890).  See also ref. 28, p.123.\\
26. Ref. 25, p.715.\\
27. See, e.g. M.J. Klein, ``The Maxwell-Boltzmann Relationship'', in: A.I.P.
Conference Proceedings, {\em{``Transport Phenomena''}}, J. Kestin, ed., 
Amer. Inst. Phys., New York (1973) p.300-307.\\
28. L. Boltzmann: ``\"{U}ber die Eigenschaften monozyklischen und andere
damit verwandter Systeme'', Zeitschr.f. R.u.Angew. Math (Crelles Journal)
{\underline{98}}, 68-94(1884); W. A. Band III,p.122 - 152.\\
29. Ref. 28,p.123,footnote 1.\\
30. Ref. 28, p.122. This was pointed out to me by G. Gallavotti, cf.
G. Gallavotti, ``Ergodicity, Ensembles and Irreversibility'', J. Stat. Phys.
 {\underline{78}}, 1571-1589 (1995).\\
31. (a) L. Boltzmann:  ``\"{U}ber einige F\"{a}lle, wo die lebendige Kraft 
nicht integrierender Nenner des Differentials der zugef\"{u}hrten Energie ist'',
Wien. Ber. {\underline{92}}, 853-875 (1885); W.A. Band III, p.153-175; (b) id.,
``Neuer Beweis eines von Helmholtz aufgestellten Theorems betreffende die
Eigenschaften monozyclischen Systeme'', G\"{o}tt. Nachr. (1886) 209-213; 
W.A. Band III, p.176-181.\\
32.  (a) See e.g., L. Boltzmann, ``\"{U}ber die Unentbehrlichkeit der
Atomistik in der Naturwissenschaften'' in {\em{Popul\"{a}re Schriften}}, 
second ed., Barth, Leipzig (1919) p. 141-157; (b) ref.22,p.216.\\
33. See ref. 28,p.134.\\
34. See ref. 10(a) p.30,footnote 83; ref.10(b) p.89,note 88.\\
35. G. D. Birkhoff, (a) ``Proof of the Ergodic Theorem'', Proc. Nat. Acad.
USA {\underline{17}}, 656-660 (1931); (b) id., ``Probability and Physical 
Systems'', Bull. Amer. Math. Soc., 361-379 (1932); (c) G. D. Birkhoff and 
B. 0. Koopman, ``Recent  Contributions to the Ergodic Theory'', Proc. Nat. 
Acad. USA {\underline{18}}, 279-287 (1932) and (d) G. D. Birkhoff and P. A. 
Smith, ``Structure Analysis of Surface Transformations'', J. Math. Pures 
Appl. {\underline{7}, 345-379 (1928).  See also: A. I. Khinchin: 
{\em{Mathematical Foundations of Statistical Mechanics}}, Dover, New York, 
Ch. II, p.19-32; E. Hopf, {\em{Ergodentheoric}}, Chelsea Publ. Cy., New York 
(1948).\\
36. J. W. Gibbs, {\em{Elementary Principles in Statistical Mechanics}},  
Yale University Press (1902); also Dover Publications, NY, (1960).\\
37. Ref 36, p.42-45.\\
38. Ref 10 (a) p.53-70; (b) p.44-63.\\
39. L. S. Ornstein: ``Toepassing der Statistische Mechanica van Gibbs op  
Molekulair-Theoretische Vraagstukken'', Leiden (1908).\\
40. See e.g. J. P. Eckmann and D. Ruelle, ``Ergodic Theory of Chaos and Strange
Attractors'', Rev. Mod. Phys. {\underline{57}} 617-656 (1985), p.639.\\
41. J. C. Maxwell: ``Tait's Thermodynamics'', Nature {\underline{17}},  
257-259 (1878); The Scientific Papers of James Clerk Maxwell, Dover Publ., 
NY (1952), Vol.2, p.660-671; see p.669.\\
42. D. J. Evans and G. P. Morriss, {\em{Statistical Mechanics of Nonequilibrium
Liquids}}, Academic Press, New York (1990) p.171.\\
43. D. J. Evans, E. G. D. Cohen and G. P. Morriss, ``Probability of Second Law
Violations in Shearing Steady States'', Phys. Rev. Lett. {\underline{71}}, 
2401-2404 (1993); id. {\underline{71}}, 3616 (1993).\\
44. G. Gallavotti and E. G. D. Cohen, ``Dynamical Ensembles in Nonequilibrium
Statistical Mechanics'', Phys. Rev. Lett. {\underline{74}}, 2694-2697 (1995);
ibid., ``Dynamical Ensembles in Stationary States'', J. Stat. Phys.
{\underline{80}}, 931-970 (1995).\\
45. D. J. Evans, E. G. D. Cohen and G. P. Morriss, ``Viscosity of a simple
fluid from its maximal Lyapunov exponents'', Phys. Rev. A{\underline{42}},  
5990-5997 (1990); See also: H. A. Posch and W. G. Hoover, ``Lyapunov 
Instability of Dense Lennard-Jones Fluids'', Phys. Rev. A{\underline{38}},
473-482 (1988); id., ``Equilibrium and Nonequilibrium Lyapunov Spectra for 
Dense Fluids and Solids'', Phys. Rev. A {\underline{39}}, 2175-2188 (1989).\\
46.  P.  Gaspard  and  G.  Nicolis,  ``Transport  Properties, Lyapunov 
Exponents and Entropy Per Unit Time'', Phys. Rev. Lett.{\underline{65}}, 
1693-1696 (1990).\\
47. (a) P. Cvitanovi\v{c}, P. Gaspard and T. Schreiber, ``Investigation of 
the Lorentz Gas in Terms of Periodic Orbits'', Chaos {\underline{2}, 85-90
(1992); (b) W. W. Vance, ``Unstable Periodic Orbits and Transport Properties
of Nonequilibrium Steady States'', Phys. Rev. Lett. {\underline{69}}, 
1356-1359 (1992).\\
48. (a) G. Morriss and L. Rondoni, ``Periodic Orbit Expansions for the Lorentz
Gas'', J. Stat. Phys.{\underline{75}}, 553-584 (1994); (b) G. P. Morriss, L. 
Rondini and E. G. D. Cohen, ``A Dynamical Partition Function for the Lorentz 
Gas'', J. Stat. Phys. (1994).\\
49. N. J. Chernov, G. L. Eyink, J. L. Lebowitz and Ya. G. Sinai, ``Derivation
of Ohm's Law in a Deterministic Mechanical Model'', Phys. Rev. Lett.
{\underline{70}} 2209-2212 (1993); id.  ``Steady State Electrical Conduction 
in the Periodic Lorentz Gas'', Comm. Math. Phys. {\underline{154}}, 569-601 
(1993).\\
50. Ref.40, p.636.\\
51. 1 am indebted for this suggestion to Prof. B. Knight of The Rockefeller
University.\\
52. See, e.g., F. E. Manuel, {\em{A Portrait of Isaac Newton}}, Harvard Univ. 
Press, Cambridge, MA (1968).\\
53. Exceptions are, e.g., M. J. Klein, {\em{``Paul Ehrenfest''}}, Vol. I, 
North-Holland, Amsterdam (1970); C. W. F. Everitt, ``Maxwell's Scientific
Creativity,'' in: {\em{``Springs of Scientific Creativity''}}, R. Aris, 
H. T. Davis, R. H. Stuewer, eds., Univ. Minnesota Press, Minneapolis, MN (1983)
Ch. 4, p.71-141; M. Dresden, {\em{``H. A. Kramers: Between Tradition and 
Revolution''}}, Springer Verlag (1987) and W. Moore, 
{\it{``Schr\"{o}dinger''}}, Cambrige University Press (1989).\\
54. H. A. Lorentz, ``Ludwig Boltzmann'', Commemoration oration at the meeting
of the German Physical Society, 17 May 1907; Verhandl. Deutsch. Physik. Gesells.
{\underline{12}}, 206-238 (1907); Collected Works, Vol. IX, p.359-390; p.389.\\
55. Ref.22, p.226,227.\\
56. See e.g. L. Boltzmann, ``Statistische Mechanik'' in ref. 22, p.358.\\
57. Ref.22, p.227: ``O unbescheidener Sterbliche! Dein Los ist die Freude am 
Anblicke des wogenden Kampfes!''.

\end{document}